\documentclass[aps,prx,twocolumn,superscriptaddress,nofootinbib]{revtex4-2}

\usepackage[utf8]{inputenc}
\usepackage[T1]{fontenc}
\usepackage{hyperref}
\usepackage{xcolor}

\usepackage[symbol]{footmisc}
\usepackage{amsmath,amsthm,amssymb}
\usepackage{graphicx} 
\usepackage{mathtools}
\usepackage{epstopdf}
\usepackage{tabularx}
\usepackage{tabularray}
\usepackage[normalem]{ulem}
\usepackage[T1]{fontenc}

\usepackage{float}

\usepackage{algorithm}
\usepackage{algpseudocode}
\usepackage{braket}
\usepackage{tikz}
\usetikzlibrary{quantikz2}

\DeclareGraphicsExtensions{.png,.pdf}

\begin{document}

\title{Squeezing-Enhanced Photon-Number Measurements for GKP State Generation}

\author{Paul Renault}
\affiliation{QC82 Inc., 7757 Baltimore Ave, College Park, MD 20740}

\author{Patrick Yard}
\affiliation{QC82 Inc., 7757 Baltimore Ave, College Park, MD 20740}

\author{Raphael Pooser}
\affiliation{QC82 Inc., 7757 Baltimore Ave, College Park, MD 20740}

\author{Hussain Zaidi}
\email{hussain@qc82.tech}
\thanks{Corresponding author}
\affiliation{QC82 Inc., 7757 Baltimore Ave, College Park, MD 20740}

\date{\today}

\begin{abstract}

We present an architecture for the generation of GKP states in which quadrature squeezing operations are used to control the average photon number statistics of probabilistic photon number measurements on Gaussian resource states. Specifically, we present an architecture employing a teleportation-based squeezing protocol and polynomial-gate applications integrated into a time-multiplexed multi-mode cluster state to generate cat states with high amplitudes, which are consequently used to generate GKP states with high quadrature effective squeezing. Compared to our previous work\cite{renault2025end}, in addition to using squeezing as a resource, the present architecture reduces damping and noise by minimizing the number of homodyne measurements required in GKP state generation. We demonstrate the effectiveness of these improvements—including dynamic input-state resetting and an improved breeding algorithm—by achieving a fault-tolerance threshold of $11.5$ dB cluster squeezing using the RHG surface code for error correction, without requiring active switching or photon-number resource states.

\end{abstract}

\maketitle

\section{Introduction}
Quantum computers aim to harness fundamental principles of quantum mechanics—most notably entanglement and superposition—to outperform classical computers \cite{Shor1994, Deutsch1992}, with recent demonstrations of quantum advantage \cite{Arute2019,Zhong2020} fueling excitement in the field. In parallel, there has been growing recognition of the need to scale quantum computers to large qubit counts and gate depths in order to tackle a broad range of practically useful problems that remain intractable for classical computers \cite{wecker2014}. 
To develop large-scale quantum computers robust to inherent noise and loss in quantum systems, the protocol of \emph{fault tolerant quantum computing} has become paramount, in which quantum information is encoded in topological arrangements of collections of physical qubits (so called \emph{logical qubits}) dictated by the chosen error correcting code such that errors arising in individual qubits can be detected through measurements and corrected in real-time \cite{Shor1995,Gottesman1997,Steane1996,Kitaev2003}. Once individual qubits are below a certain loss threshold, increasing the size of the physical qubit encoding— often quantified by the code distance— leads to \emph{lower} computational error rates.  

Many quantum modalities are being pursued to create large-scale fault tolerant quantum computers, including neutral atoms \cite{xu2024constant, sales2025experimental}, superconducting qubits \cite{google2025quantum,putterman2025hardware}, photonic polarization and space encoding \cite{psiquantum2025manufacturable, Knill2001,Kok2005}, and continuous degree photonic quadratures \cite{aghaee2025scaling}. Notable advantages of photonic quantum computing over other modalities are the abilities to apply quantum gates at very high rates, modular construction without transduction, time-multiplexing of the quantum signal to reduce physical footprint, and arbitrary topological arrangements of qubits leading to a variety of error correcting codes that may be hard to achieve in other platforms. Photonic continuous-variable quantum computing (CVQC) \cite{Lloyd1999,Fabre2020,Pfister2019, Menicucci2006}– in which information is encoded in the quantum mechanical quadratures of the photons - is further promising due to deterministic entanglement generation between qubits using passive photonic elements \cite{Menicucci2010} and near-deterministic ways of creating photonic qubits \cite{Walschaers2021}. Gottesman Kitaev Preskill (GKP) qubits \cite{Gottesman2001} are a popular choice in CVQC as the entire Clifford+T universal gate set can be implemented with only homodyne measurements on GKP states and related magic states, which can be generated from GKP states \cite{Baragiola2019, Bravyi2005, renault2025end}. Hence, significant work has focused on feasible methods for generating GKP qubits with high \emph{effective squeezing} (a figure of merit bounding the measurement error on an individual GKP qubit \cite{Weigand2018}). Known methods to create GKP qubits with a high probability require active switching \cite{aghaee2025scaling}, high photon number resolution \cite{tzitrin2020progress,aghaee2025scaling}, high Gaussian squeezing \cite{takase2021generation}, or photon number states \cite{Winnel2024}, which are challenging for large-scale fault tolerant quantum computing. 

Our earlier work \cite{renault2025end} proposed the use of time-frequency multiplexing, passive linear optics, low photon number resolution and homodyne measurements to create GKP states nearly-deterministically and high-fidelity magic states with a high success probability. The physical implementation relied on engineering a multi-mode frequency-time multiplexed cluster state to implement the Photon-number Assisted Node Teleportation Method (PhANTM) \cite{Eaton2022Phantm} to create cat states \cite{Walschaers2021} on a subset of the nodes of the cluster state. Mixtures of cat states and momentum squeezed states were adaptively bred together with linear optics to generate GKP states, while magic states were created from GKP and cat states using homodyne measurements. Fault tolerance thresholds were found using the RHG surface code implemented in a modular architecture of interconnected GKP qubits. In this paper, we describe a new protocol with dynamic Gaussian squeezing applied between PhANTM steps on a photonic time-multiplexed two-mode squeezed cluster state to generate higher-quality GKP states for fault tolerant quantum computing with a lower squeezing threshold. In the final analysis, we achieve a fault tolerance threshold of $11.5$ dB with $10$ steps of PhANTM using the RHG surface code for error correction, which, to the best of our knowledge, is state-of-the-art for a lossless architecture using Gaussian resource states. The passive protocols are implemented on a cluster state engineered to minimize the application of noisy homodyne measurements. The cluster state is initialized dynamically to enhance the number of photons subtracted. We employ an adaptive breeding scheme in which cat states are combined with momentum-squeezed states in a controlled mixture to generate GKP states. While our end-to-end architecture creates GKP states with stochastic effective squeezing in the two CV quadratures, the effect of this stochasticity on the fault tolerance threshold is shown to be minimal, which implies that strictly targeting GKP states with a desired effective squeezing is likely unnecessary for fault-tolerant photonic quantum computers in the intermediate-term horizon.

The article is organized as follows. We start in section \ref{sec:conventions} with a brief overview of the conventions used in CVQC and provide an introduction to PhANTM — a probabilistic method to generate non-Gaussian states from Gaussian squeezed states. In section \ref{sec:antisqueezing}, we motivate the usefulness of squeezing in combination with PhANTM to create highly non-Gaussian output states, specifically, cat states. Then in section \ref{sec:architecture} we present the complete simulation of creating GKP states in the presence of noise starting from resource two-mode squeezed states.  Finally, in section \ref{sec:qec} we estimate the fault tolerance threshold using the surface code on a macronode RHG lattice and compare it to the state of the art. 

\section{Conventions and overview of the architecture}
\label{sec:conventions}
CV quantum systems can be described by quadrature operators, $q$ and $p$, such that they obey the commutation relation $[q,p]=i$. These operators are combinations of annihilation ($a$) and creation ($a^\dagger$) operators,
%\begin{equation}
    $q=\frac{a^\dagger+a}{\sqrt{2}}$ and 
%\end{equation}
%\begin{equation}
    $p=\frac{a-a^\dagger}{i\sqrt{2}}$
% \end{equation}
with $[a,a^\dagger]=1$. In photonics, $q$ and $p$ refer to the electromagnetic field
quadrature operators while $a$ and $a^\dagger$ are, respectively, the photon annihilation and creation operators. To describe and visualize a CV quantum state $\rho$, we often use the Wigner function, which is a quasi-probability distribution defined as
\begin{equation}
    W(q,p)[\rho]=\frac{1}{\pi}\int_{-\infty}^{+\infty}\mathrm{d}y e^{2ipy}\braket{q-y|\rho|q+y}.
\end{equation}

Below, we list the single mode and two-mode operators used in the description of our architecture. We define the displacement operator as
$D(\alpha)=e^{\alpha a^\dagger - \alpha^* a}$, where $\alpha$ is a complex number representing the displacement of the Wigner distribution in phase space. The squeezing operation is defined as $S(r)=e^{\frac{1}{2}r^*a^2-ra^{\dagger2}}$ where $r=|r|e^{i\theta_r}$. This operation squeezes the Wigner distribution by $|r|$ along an axis rotated in the phase space by $\theta_r$. The final single-mode operator we define is the rotation operator, $R(\theta)=e^{-i\theta a^\dagger a}$, where $\theta$ is the angle of rotation of the Wigner distribution in the phase space.
To create entanglement, we need multi-mode operations, of which the beam splitter operation is the most experimentally
accessible. This operation is defined as $B_{1,2} = e^{\theta_{1,2}(a_1^\dagger a_2e^{i\phi}-a_1a_2^\dagger e^{-i\phi})}$, where $\theta_{1,2}$ is the beam splitter angle that determines mode coupling between the two input modes and $\phi$ is the phase. The controlled-Z operation is another two-mode entangling operation defined as $C_Z(g)=e^{g q_1 q_2}$, where $g$ is the weight of the $C_Z$ interaction between the two modes. Experimentally, this operation can be performed using two-mode squeezing such that if $r_0$ is the squeezing parameter of the source, the weight of $C_Z$ is then $g=\tanh{2r_0}$. \cite{EatonThesis2022,Eaton2022Phantm}. 

Our architecture relies on creating squeezed photonic cat states which are superpositions of displaced squeezed vacuum states. These states are parametrized by their amplitude $\alpha$ and their squeezing $r'$, described as $\ket{\pm\alpha,r'}\approx \left(D(\alpha) \pm D(-\alpha)\right)S(r')\ket{0}$, where $\ket{0}$ is the zero photon number vacuum state, and the sign of the displacement determines the parity of the state. In order to keep the analysis simple, we define the parameter $\alpha_c$ (\emph{the corrected amplitude}) as the amplitude of the cat state when it is corrected to zero squeezing. For $\alpha$ and $r'$ real numbers, we have $\alpha_c=\alpha\left(\cosh(r')+e^{i\theta'}\sinh(r')\right)$, where $\theta'$ is the orientation of the squeezing operation, which we set to zero. 

Mixtures of squeezed cat states and squeezed vacuum states are used to generate our photonic
GKP states, which are the $+1$ eigenstate of displacement operators $D_q(\sqrt{2\pi})$ and $D_p(\sqrt{\pi})$, generated from cat states using adaptive breeding protocols \cite{Weigand2018,renault2025end}. Ideal (infinite-energy) GKP $\ket{0_L}$ and $\ket{1_L}$ qubits can be defined as \cite{Gottesman2001}
\begin{equation}
    \ket{0_L} = \sum_{n\in \mathbb{Z}} \ket{2n\sqrt{\pi}}
\end{equation}
\begin{equation}
    \ket{1_L} = \sum_{n\in \mathbb{Z}} \ket{(2n+1)\sqrt{\pi}},
\end{equation}
while realistic GKP states are approximated as a superposition of squeezed vacuum states displaced by integer multiples of $\sqrt{\pi}$. The peak width indicates how closely the states approximate the ideal case, often quantified by the \emph{effective squeezing}, $\Delta_D = \frac{1}{u}\sqrt{\ln\left(Tr(D(u)\rho)\right)}$, where $u$ defines the displacement, $\rho$ is the density matrix and $Tr$ is the trace operator \cite{Weigand2018}. 

\begin{figure}
    \centering
    \includegraphics[width=0.4\textwidth]{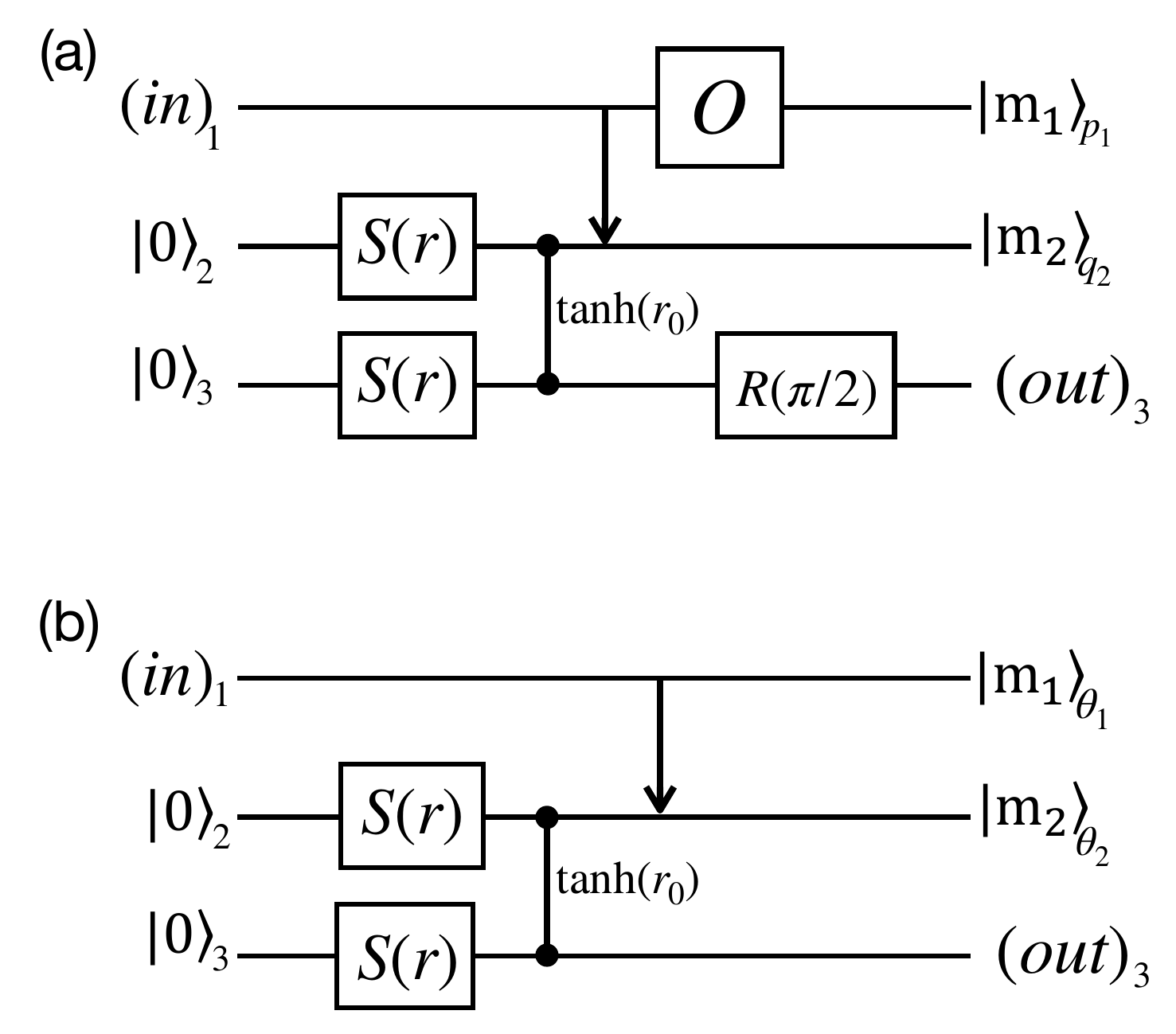} 
    \caption{(a) Circuit for PhANTM on a dual-rail quantum wire. $O$ stands for multiple photon subtractions, i.e., $O$ represents multiple $\mathcal{O}$ in Eq. \ref{eq:subtraction} combined. (b) Circuit for squeezing gate using dual-rail quantum wire which resembles the PhANTM circuit, with key differences being the removal of photon subtraction and the change of the homodyne detection angles depending on the desired squeezing. In both circuits, arrow represents a beam splitter and line with circular ends represents a $C_Z$ interaction.}
    \label{fig:circuits}
\end{figure}

One of the fundamental building blocks of our architecture is PhANTM, schematically shown in Fig. \ref{fig:circuits} for the dual-rail quantum wire \cite{Eaton2022Phantm} in which multiple photon subtractions (represented by the operator $O$) followed by homodyne measurements are repeatedly enacted to apply polynomial operators to the input quantum state and build up non-Gaussian cat states. Deterministic photon subtraction can be
modeled with integer powers of the annihilation
operator applied to the input state. In reality, photon subtraction is a probabilistic process implemented with a
beam splitter along with a photon number resolving (PNR) detector
on the reflected arm \cite{ourjoumtsev2006generating,Ourjoumtsev2007,Wakui2007} or by employing non-linear optics \cite{Ra2020} and a PNR detector. The operator to model photon subtraction with a beam splitter can be expressed as:
\begin{equation}
    \mathcal{O}_n = \frac{(-1)^n e^{n\beta}/2}{\sqrt{n!}}\left(2\sinh(\beta)\right)^{n/2}a^n N(\beta)
    \label{eq:subtraction}
\end{equation}
where $\beta = -\ln{t}$, $t$ the transmittance of the beam splitter, $n$ is the number of photons subtracted, and $N(\beta)=e^{-\beta a^\dagger a}$ is a damping operator dependent on the transmittance chosen for the beam splitter.
\\

The final component we review here is the teleportation based squeezing gate represented in Fig. \ref{fig:circuits} for a dual-rail quantum wire \cite{Alexander2014}, in which squeezing is applied to the input state by appropriately setting the homodyne measurement angles. Taking into account the noise introduced because of finite squeezing, the gate can be 
expressed as:
\begin{equation}   
    \mathcal{S}(r, r_a) = \mathcal{N}(r)C_m(m_1,m_2)S(r_a)
    \label{eq:Sgate}
\end{equation}

with   
\begin{equation}
    \mathcal{N}(r)=\exp(-\epsilon q^2/2)\exp(-\epsilon p^2/2\tanh(2r_0)^2)
    \label{eq:noise}
\end{equation}
and where $C_m$ is the displacement induced by non zero homodyne measurements (the expression of $C_m$ can be found in \cite{Alexander2016a}). In \ref{eq:Sgate}, $r_a$ is the squeezing parameter determined by the homodyne basis and  $\epsilon = \mathrm{sech}(2r_0)$, where $r_0$ is the squeezing parameter of a two-mode squeezing source. For concreteness, the cluster squeezing $r$ can be expressed in terms of $r_0$ as
\begin{equation}
    r=\ln(\sqrt{\epsilon})=\ln(\sqrt{\mathrm{sech}(2r_0)}).
\end{equation} 

Throughout the manuscript, when noise damping is mentioned, we refer to the operator $\mathcal{N}$, which differs from the operator $N$ defined in Eq. \ref{eq:subtraction}. 

\section{Squeezing-Enhanced Photon Number Measurements}

\label{sec:antisqueezing}
\begin{figure*}
    \centering
    \includegraphics[width=0.95\textwidth]{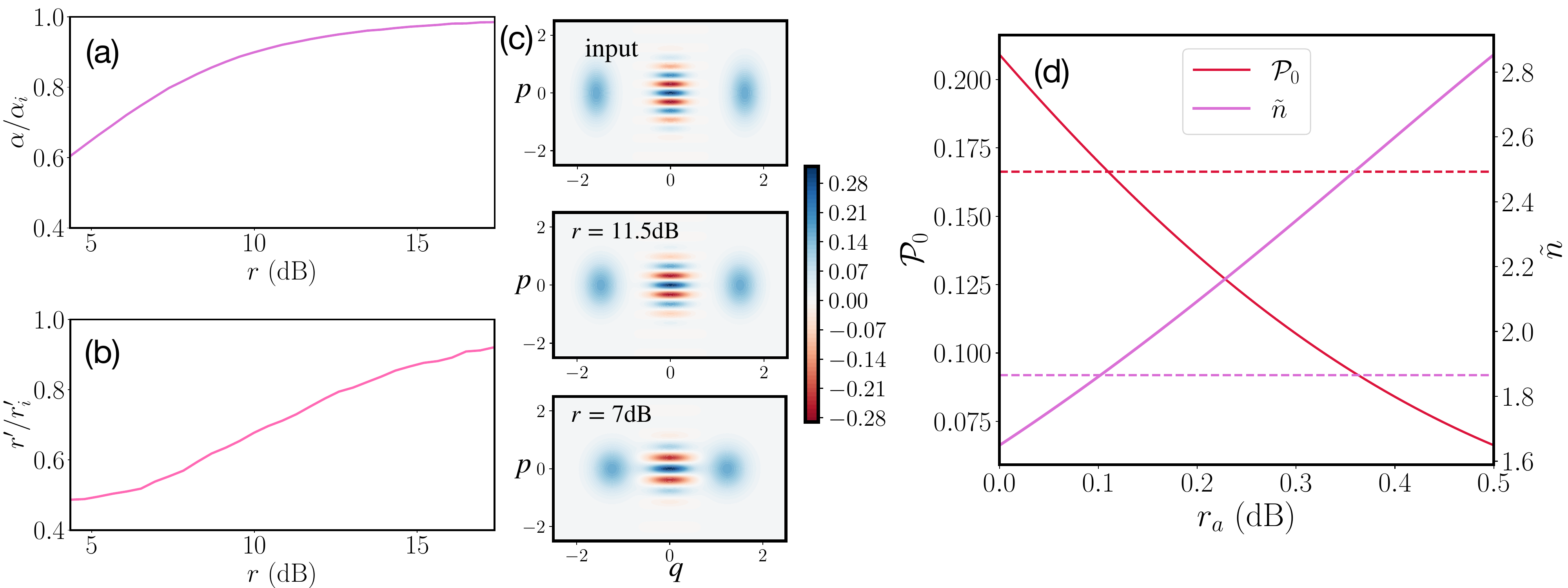} 
    \caption{Effect of noise on cat states. a) ratio of cat state size, between after and before noise channel $\mathcal{N}$, as a function of the cluster squeezing $r$ of the channel. b) ratio of cat state squeezing as a function of the cluster squeezing $r$. c) Wigner function of cat states. An input state is taken with a size of $\alpha_i=2$ and a squeezing of $r^\prime_{i}=0.5$. The output state of the noise channel are shown with $r=11.5$ dB and $r=7$ dB. d) $\tilde{n}$ (left axis) and $\mathcal{P}_0$ (right axis) as a function the anti-squeezing parameter $r_a$ where the anti-squeezing gate is applied before PhANTM (solid line). We assume cat state input with $\alpha_i= 3$,  $r^\prime=0.5$, and squeezing in the cluster state as well as the anti-squeezing gate to be $r=11.5$ dB. For these simulations, we consider one photon subtraction in a PhANTM step. For reference, we show $\mathcal{P}_0$ and $\tilde{n}$ in the case where PhANTM is applied without an anti-squeezing gate (dashed line). Subscript $i$ denotes input or initial parameter.}
    \label{fig:noise_channel}
\end{figure*}

We start by motivating why squeezing is helpful in controlling photon number statistics in combination with PhANTM for the goal of efficiently generating cat states. A first order goal in creating large cat states is to subtract a large number of photons from a squeezed momentum or cat state, leading to applications of higher degree polynomial gates to the input state and, consequently, higher amplitude of the output state \cite{Eaton2022Phantm}. 

In PhANTM, the probability of photon subtraction can be controlled by changing the transmittance of the photon subtraction setup, but higher transmittance introduces higher damping of the state, as can be seen in eq. \ref{eq:subtraction}. This leads to an optimization between maximizing the probability of photon subtraction and minimizing the transmittance to maximize the average corrected amplitude of cat states. 

The average number of photons in a cat state with amplitude $\alpha$ is $\alpha^2e^{-2r_a}+\sinh^2{r_a}$, where $r_a$ is the squeezing applied to the cat state. For large enough cat states ($\alpha\geq1$) and moderate squeezing values of $r_a\approx 1$, the average photon number is significantly less than that of a zero-squeezed cat state. Since squeezing (or anti-squeezing) a cat state changes the photon distribution of a cat state, one can also use anti-squeezing to change the probability of photon subtraction. This dynamic control of the photon subtraction probability does not change the damping angle but changes the input density matrix instead. As shown later in \ref{sec:phantm}, squeezing can be controlled in an integrated time-frequency multiplexed cluster state with homodyne measurement angles without dynamically changing the integrated photonic circuit or introducing variable beam splitters. In the following, we quantitatively describe the effect of squeezing with PhANTM considering realistic cluster squeezing levels and moderate cat state amplitudes.

First, consider that while time multiplexing enables the near-deterministic generation of cat states with repeated applications of PhANTM, this comes at the cost of introducing noise (damping) because of the finite squeezing of the ancilla state (modes 2 and 3 in Fig. \ref{fig:circuits}). Fig. \ref{fig:noise_channel} shows cat state parameters before and after passing through a noise channel $\mathcal{N}$ (see \ref{eq:noise}), which is modeled here as a teleportation gate on the dual-rail quantum wire. We observe that while the cat size $\alpha$ can be maintained at a high cluster squeezing of around $15$ dB for a moderate cat input size of $\alpha_i=2$, the squeezing $r'$ is not readily preserved. One way to mitigate noise and damping to increase the probability of photon number measurements is by using an anti-squeezing gate. Fig. \ref{fig:noise_channel} (d) shows the results of using a teleportation-based anti-squeezing gate as defined in eq. \ref{eq:Sgate}. We define $\mathcal{P}_0$ and $\tilde{n}$ as the probability of subtracting zero photons  and the expectation number of subtracted photons, respectively.
We find that applying an anti-squeezing gate before PhANTM simultaneously reduces $\mathcal{P}_0$ and increases $\tilde{n}$ as $r_a$ grows for input cat states with $\alpha_i=3$  and $r^\prime_i=0.5$. At $r_a=0$, however, the anti-squeezing operation reduces to a noisy teleportation channel, which increases the probability of measuring zero photons compared to the case without any supplementary gate (dashed horizontal line). As $r_a$ grows, the additional noise from teleportation-based anti-squeezing gate is progressively compensated, and the advantage of the anti-squeezing gate becomes apparent. This crossover is visible in Fig. \ref{fig:noise_channel}, where the solid and dashed curves intersect near $r_a\approx0.1$ for both $\mathcal{P}_0$ and $\tilde{n}$, after which larger values of $r_a$ continue to improve the probability of measuring several photons in PhANTM. 

\section{Integrated architecture}
\label{sec:architecture}
After motivating squeezing as a control mechanism in generating non-Gaussian states, this section focuses on developing an integrated architecture that uses squeezing with PhANTM to create cat states, and adaptive breeding to create GKP states from cat states. We show that with targeted cluster state engineering, squeezing and PhANTM can be applied without active switching and the input momentum squeezed resource states can be actively reset to further improve cat state generation. One byproduct of the new protocols is that the number of teleportations needed in the full GKP generation pipeline is minimized compared to \cite{renault2025end}. 

\subsection{Cat state generation}
\label{sec:phantm}

PhANTM and squeezing can be integrated with a time-frequency-multiplexed cluster state in many ways, but our goal is to present methods that enable their integration without active switching while minimizing the number of teleportations that contribute to the damping of the input quantum state (shown, e.g., in the noise factor $\mathcal{N}$ in equation \ref{eq:Sgate}). The dual-rail time-frequency cluster state, tailored to enable repeated application of teleportation-based squeezing and PhANTM protocols over multiple time steps, is presented in Fig. \ref{fig:cat_generation}.a (see the Supplemental Material Sec. \ref{sec:appB} for engineering such a cluster state starting with a two-mode squeezed resource state, e.g., from a resonant cavity \cite{Yi70modes2023,Yang2021}). On this cluster state, PhANTM and squeezing operations are applied sequentially in time ($\tau_i$, $\tau_{i+1}$, ...) (see Fig.\ref{fig:cat_generation}.a). The separation between two time steps is determined by the physical constraints of the experiment, such as the repetition rate of the squeezing source or the frequency bandwidth of homodyne detectors or PNR detectors. Each time step requires access to two spatial modes (two propagating waveguides) and two frequency modes (resonant modes within the cavity generating the squeezing). PhANTM is applied to one frequency mode, $\nu_1$, with $\nu_{2a}$ as the output mode, while the squeezing gate is applied to the other frequency mode, $\nu_2$, yielding $\nu_{1b}$ as the output mode. Compared to \cite{renault2025end}, this approach requires no homodyne measurements for redirecting cluster entanglement or for deleting modes from the cluster, which conserves the squeezing of the cluster state. As a consequence of the new cluster state construction, the beam splitter phase in the PhANTM circuit is changed from $\pi/2$ to $0$ as shown in Fig.\ref{fig:cat_generation}.a.

This architecture allows us to implement a reset mechanism, which we motivate and describe as follows. Since the initial input state for cat generation is a squeezed vacuum state, if no photons are subtracted during the first step, the output remains Gaussian, albeit with reduced squeezing. This reduction, in turn, lowers the probability of photon subtraction in the following PhANTM iteration, potentially leading to a series of unsuccessful subtraction attempts. To prevent this cascade, we introduce a reset mechanism: if no photon is subtracted during a given PhANTM step, the input state is reinitialized to the squeezed cluster state before proceeding to the next iteration by performing homodyne detection in the $q$ quadrature basis. In this way, the initial squeezed vacuum is reused until a nonzero photon measurement occurs. 

Our strategy for optimizing the amplitude and the squeezing of the final cat state proceeds as follows: after each PhANTM step, an anti-squeezing gate $\mathcal{S}(r,r_a^{(1)})$ is applied. Once the total number of subtracted photons reaches a predefined threshold $\mathcal{T}_{\mathrm{ph}}$, the anti-squeezing strength is reduced by switching to a smaller parameter $r_a^{(2)} < r_a^{(1)}$. The values of $r_a$ were obtained after empirical optimization to maximize $\alpha_c$ averaged over the Monte Carlo runs. For each simulation performed for a given value of $r$, the anti-squeezing gate parameters are the following: $r_a^{(1)}=2.39$ dB, $r_a^{(2)}=0.43$ dB and $\mathcal{T}_{\mathrm{ph}}=55$. To keep the simulation runtime feasible, and as the displacement can be corrected, we assume homodyne detection results to be zero and discard $C_m$. For the simulations, we picked relatively small values of $r_a^{(1)}$ and $r_a^{(2)}$ given the constraints of keeping the Fock dimension reasonable for a feasible runtime of our simulations. In reality, we do not see any reason to avoid higher values of anti-squeezing to further improve the fault tolerance threshold in the lossless case (see Fig.\ref{fig:noise_channel}.d). True optimization of the squeezing at each PhANTM step to control cat generation will require more extensive Monte Carlo simulations. 
\begin{figure*}
    \centering
    \includegraphics[width=1\textwidth]{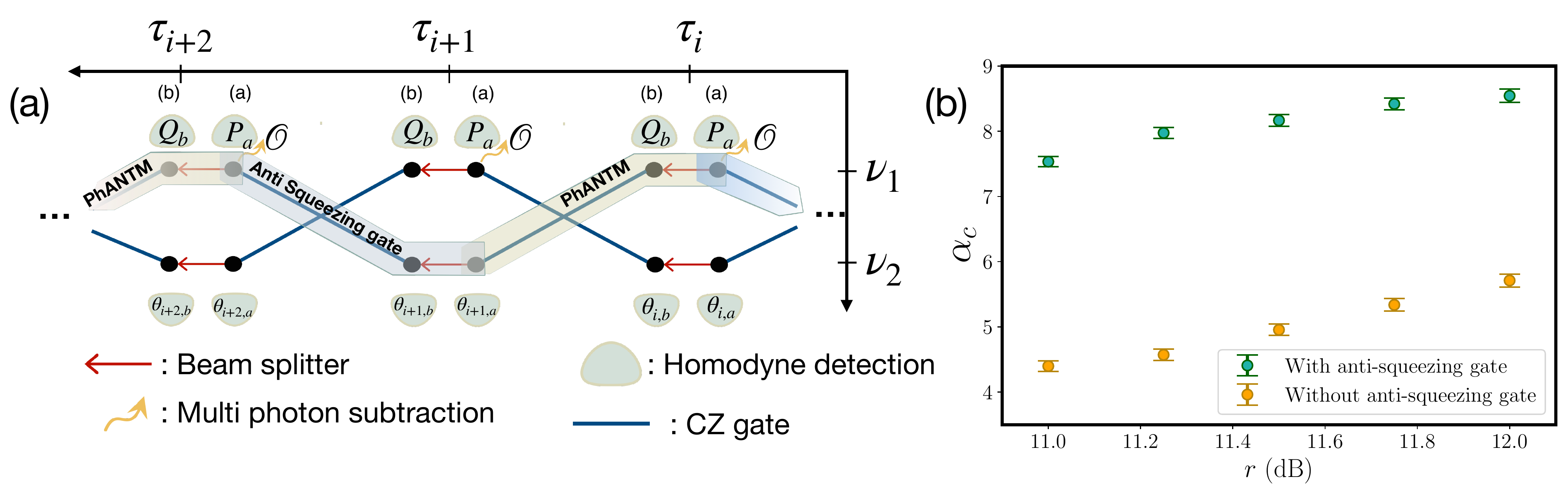} 
    \caption{Cat state generation protocol and results.(a) PhANTM and squeezing are applied successively in a dual rail frequency-time cluster state. In PhANTM, homodyne bases are $p$ and $q$ while for squeezing, the angle $\theta_{i,a}$ and $\theta_{i,b}$ are chosen dynamically (see \cite{Alexander2016a}) depending on how much the cat state needs to be squeezed (see eq.\ref{eq:Sgate}). (b) Result from PhANTM simulation: $\alpha_c$ (corrected amplitude) as a function of cluster squeezing $r$. Each dot is the mean of a Monte Carlo simulation with 1000 trials, while error bars show the standard deviation of the mean. For baseline comparison without anti-squeezing gate, PhANTM is applied both on mode $\nu_1$ and $\nu_2$.}
    \label{fig:cat_generation}
\end{figure*}
In our architecture, we assume eight PNR detectors for every quantum mode, each PNR detector capable of resolving only a small number of less than 10 photons. This choice is the result of a few numerical optimizations motivated by the trade-off in the number of detectors and the number of time iterations to create GKP states above the fault tolerance threshold. Regardless of the anti-squeezing applied to the signal, the damping from each photon subtraction adversely affects the quantum signal. To mitigate the reduction in photons subtracted caused by the damping, the most direct strategy is to adjust the parameters of the linear optical elements, in particular, the reflectivity of the beam splitters in PhANTM. However, this optimization is fundamentally constrained by the probabilistic nature of photon subtraction. In \cite{renault2025end}, a linear gradient was applied across the eight successive PhANTM steps. In this work, we employ a non-uniform exponential gradient defined as $\theta_{x+1} = \theta_x + \nabla(x)$, where $\nabla(x) = a\mathrm{e}^bx$, $a$ and $b$ are tunable parameters learnable offline through calibration and validation of the quantum system, and $x$ is the position of the photon subtraction within the PhANTM step. This formulation provides greater flexibility in shaping the subtraction probabilities across the PhANTM sequence. For the simulations presented in this work, we used $a=0.75$, $b=0.35$ and an initial angle $\theta_{x=0}=18^{\circ}$. These parameters were optimized to maximize the average value of $\alpha_c$ in Monte Carlo simulation of PhANTM. More simulation details are given in the supplementary section (see \ref{sec:appA}). 

Fig. \ref{fig:cat_generation}.b shows $\alpha_c$  averaged over ten iterations of PhANTM and squeezing as a function of the cluster squeezing $r$, which is also the input squeezing in the PhANTM circuit. Two scenarios are investigated: one in which anti-squeezing gate is applied following the method described in \ref{sec:antisqueezing}, and one without it, keeping the same numerical parameters as above.  We can observe that the anti-squeezing gate optimized for the numerical parameters of our architecture provides a clear advantage in cat state generation. After optimization, $\alpha_c$ increases by at least 2 units depending on $r$. This improvement directly enhances the effective squeezing of the resulting GKP states obtained through subsequent adaptive breeding. A resolution of fewer than ten photons is required for the photon detectors (see Fig. \ref{fig:photon_dist}). This photon number resolution is feasible at low temperatures with superconducting nanowire single-photon detectors \cite{ Goltsman2001,marsili2013detecting} or segmented detectors based on room-temperature semiconductor technology \cite{nehra2020photon}. 

\begin{figure}
    \centering
    \includegraphics[width=0.4\textwidth]{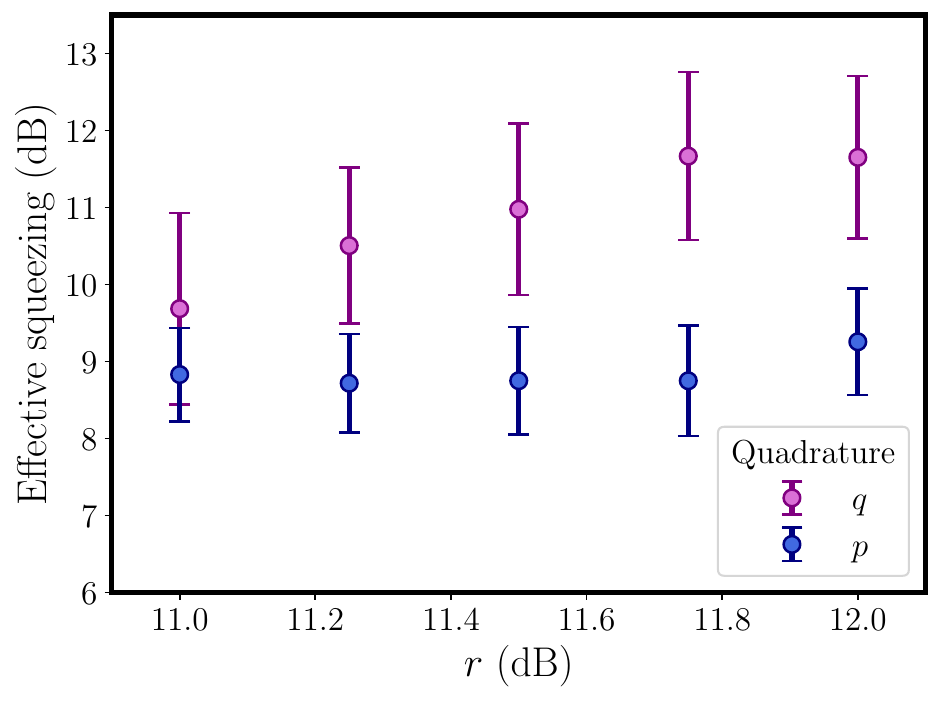} 
    \caption{GKP effective squeezing as a function of $r$ from adaptive breeding simulations. The circle shows the average over 1000 Monte Carlo runs, while the bars show the standard deviation. Simulations are performed with a Fock dimension of 65.}
    \label{fig:gkp_result}
\end{figure}
\subsection{GKP generation with adaptive breeding}\label{sec:breeding}
GKP states are generated through adaptive breeding of cat states and momentum squeezed states. Standard breeding consists of interfering two cat states on a beam splitter followed by homodyne measurement of one output mode in the $p$ quadrature \cite{Weigand2018} to create a non-Gaussian state in the unmeasured mode. By appropriately setting the amplitude of the input cat states, this procedure can yield an approximate GKP state. The required amplitude $\alpha_b$ of the input cat states depends on the number of breeding rounds $M$, given by $\alpha_b = 2^{(M-3)/2}$ \cite{Weigand2018}.

We employ an adaptive breeding protocol well-suited for cat inputs with probabilistic amplitude and squeezing, in which we breed a mixture of squeezed momentum and cat states, optimizing the squeezing of the input cat states to maximize the average effective squeezing of the resulting GKP states. The setup is schematically shown in Fig. \ref{fig:Breedingcircuit}. Each incoming cat state—characterized by its amplitude $\alpha$, its squeezing $r^\prime$ and its parity $P$— can undergo a teleportation-based squeezing gate to rescale the state to amplitude $\alpha_b$ or $2\alpha_b$. Correspondingly, we define two squeezing levels, $r^\prime_{\alpha_b}$ and $r^\prime_{2\alpha_b}$ as the squeezing of the cat state after it has been rescaled to amplitude $\alpha_b$ or $2\alpha_b$. The need for adaptive breeding arises from the fact that the squeezing in the GKP quadratures is a function of the initial squeezing of the cat states, their amplitudes, and the number of momentum squeezed states included in the input mixture. In one instance of adaptive breeding, $r^\prime_{\alpha_b}$ and $r^\prime_{2\alpha_b}$ are compared to predefined lower bounds $r^\prime_{\alpha_{\text{lb}}}$ and $r^\prime_{2\alpha_{\text{lb}}}$, which are learned from adaptive breeding runs and set the minimum acceptable squeezing levels for cat states. 
If $r^\prime_{\alpha_b} > r^\prime_{2\alpha_b}$ and $r^\prime_{\alpha_b} > r^\prime_{\alpha_{\mathrm{lb}}}$, the state is squeezed to match the amplitude $\alpha_b$. If this condition is not met, but $r^\prime_{2\alpha_b} > r^\prime_{2\alpha_{\mathrm{lb}}}$, the state is instead squeezed to reach $2\alpha_b$. If neither condition holds or if the available squeezing within the cluster state is insufficient to perform the required transformations, the cat state is replaced by a squeezed vacuum state with squeezing $r$ (the cluster squeezing parameter). Notice that these thresholds are learnable offline, and the optimization of GKP effective squeezing is multivariate, owing to the different thresholds that can be defined for cat states squeezed to even-integer multiples of $\alpha_b$. We have neglected the higher effective squeezing that can be derived from such higher order optimizations in this work. 

Applying the squeezing gate to adjust the size of the input cat states introduces an additional source of noise. We account for the noise introduced by this channel by applying the operator $\mathcal{N}(r)$ after the full PhANTM stage, which effectively reduces the squeezing of the cat states compared to the ideal case of infinite cluster squeezing (see supplementary information \ref{sec:appA}).  
We perform three adaptive breeding rounds, requiring a total of eight output states from the PhANTM protocol described in Sec.~\ref{sec:phantm}. In the simulations, the homodyne measurement outcomes are set to zero for computational speed. Note that breeding without post-selection is known to produce GKP states with higher effective squeezing \cite{Weigand2018}. In experimental implementations of adaptive breeding, non-zero outcomes require applying a displacement to the resulting GKP state at the very end of breeding to compensate for the measurement-induced displacement, as may be needed for quantum error correction. Additionally, a displacement on the $p$ axis can be applied to some of the input cat states to ensure parity matching between the interfering states. 

Table \ref{tab:lowerbound} shows the learned lower bounds for each cluster squeezing setting over 1000 Monte Carlo trials of adaptive breeding. To select these parameters, an idealized fault tolerance threshold curve was considered in the quadrature phase space. The effective squeezing average was calculated for each quadrature and the smallest/largest distance from the fault tolerance threshold curve was calculated depending on whether the average GKP state was below/above the fault tolerance threshold, respectively. The optimal lower bound squeezing values listed in the table are those that produced GKP states with the lowest error rates as measured by the distance from the threshold curve.

Fig.~\ref{fig:gkp_result} shows the effective squeezing as a function of the cluster squeezing $r$. Each data point corresponds to a simulation run with adjusted values of $r_{\alpha lb}$ and $r_{2\alpha lb}$ as gathered in Table ~\ref{tab:lowerbound}. Note that these are static parameters that are optimized for the system once. As expected, higher cluster squeezing results in higher effective squeezing. This trend is consistent with previous observations: stronger cluster squeezing yields larger cat states after PhANTM, which, when rescaled to a fixed amplitude $\alpha_b$ or $2\alpha_b$, leads to more strongly squeezed states as input to the breeding algorithm. 
\begin{table}[t]
    \caption{$\alpha$ and $2\alpha$ squeezing lower bounds used for each cluster-state squeezing $r$.}
    \label{tab:lowerbound}
    \begin{ruledtabular}
    \begin{tabular}{ccccccccc}
         $r$(dB)& $11$ & $11.25$ & $11.5$ & $11.75$ & $12$ & $12.25$ & $12.5$ \\
        \hline
        $r_{\alpha_{\mathrm{lb}}}$(dB) & 6.08 & 5.21 & 4.34 & 0.45 & 3.91 & 4.78 & 4.78\\
        $r_{2\alpha_{\mathrm{lb}}}$(dB) & 7.82 & 6.95 & 6.08 & 5.65 & 6.08 & 6.51 & 6.51\\
        
    \end{tabular}
    \end{ruledtabular}
\end{table}

\section{QEC threshold evaluation}
\label{sec:qec}
While GKP states can be used in a variety of quantum  applications \cite{zhuang2020distributed, koudia2025crosstalk, brady2024advances}, we focus on fault tolerant quantum computing in this work. To validate the effectiveness of our cat and GKP generation protocols we describe our fault tolerance simulations and compare the results of the fault tolerance threshold to state of the art \cite{renault2025end,tzitrin2021fault,takase2021generation,aghaee2025scaling}.

\subsection{RHG surface code and decoding the errors}

In order to suppress errors in quantum computation, any qubit level errors that arise from decoding the GKP code can be mitigated by concatenating the GKP code with an outer qubit level code for quantum error correction \cite{Gottesman2001,Fukui2018, tzitrin2021fault, larsen2021fault}. Many codes have been recently devised for quantum computation \cite{Calderbank1997,Steane1996,Raussendorf2007,Raussendorf2007topological}. Beyond their error-correcting performance, many modern approaches focus on minimizing resource overheads by moving to low density parity check codes, where the number of logical qubits scales with the number of physical qubits \cite{breuckmann2021,hastings2021}. 
For direct comparison with our earlier work \cite{renault2025end}, we employ the measurement based version of the surface code implemented with the macronode construction of the Raussendorf-Harrington-Goyal (RHG) lattice \cite{Raussendorf2007,Raussendorf2007topological,tzitrin2021fault}. 
This approach exploits the fact that, in continuous-variable photonic platforms, large cluster states can be generated straightforwardly from multi-mode squeezed states combined with linear optics. In this setting, the controlled-Z links of the canonical lattice are effectively replaced by GKP Bell pairs, which can be created by interfering GKP sensor states on a beam splitter. Each macronode in the RHG lattice comprises of four GKP states originating from four distinct Bell pairs. Among these, one mode is chosen to represent the node in the underlying lattice, while the remaining three ancillary modes are measured in the $q$ basis. More details about the construction of the macronode RHG lattice with passive linear optics can be found in \cite{renault2025end,tzitrin2021fault}.

\begin{figure}
    \centering
    \includegraphics[width=0.45\textwidth]{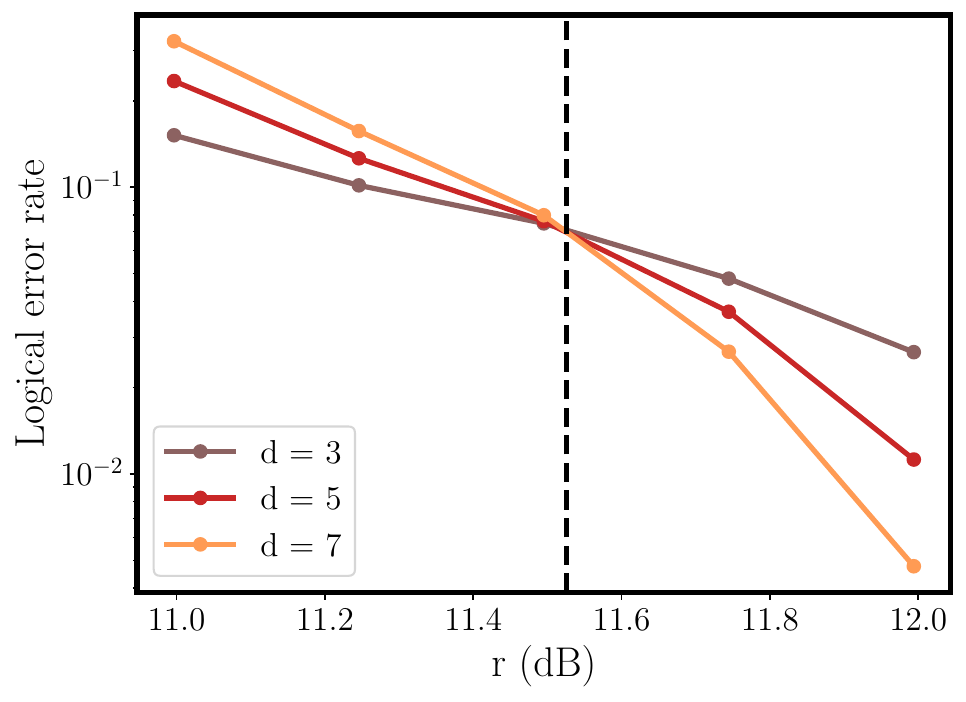} 
    \caption{Logical error rate as a function of the cluster squeezing $r$ for different code distances. The dashed line indicate the position of the quantum error correction threshold. Each dot is determined from $10^5$ points and by using MWPM decoding method \cite{Dennis2001}.}
    \label{fig:threshold}
\end{figure}

\begin{figure}
    \centering
    \includegraphics[width=0.45\textwidth]{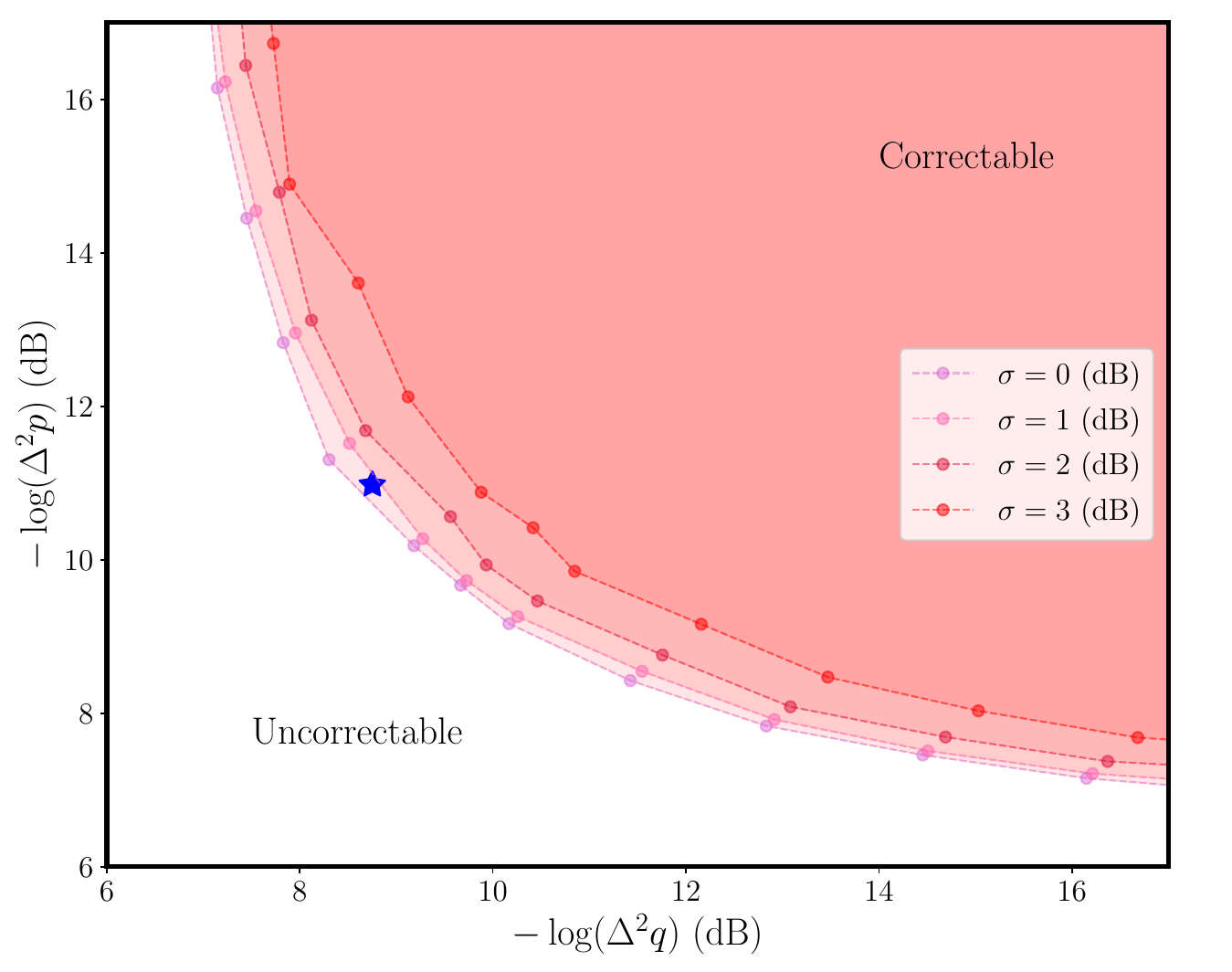} 
    \caption{2D threshold in term of the effective squeezing values in the $q$ and $p$ quadratures. The effective squeezing values are taken as the mean of normal distributions, each characterized by a corresponding standard deviation $\sigma$. These distributions are then used as input for the QEC simulations, which determine the logical error rate as function of the code distance and, in turn, the threshold. Boundaries between correctable and uncorrectable region can be defined for different value of $\sigma=0, 1, 2, 3$. The blue star represents the mean effective squeezing shown in \ref{fig:gkp_result} at $r=11.5$ dB.}
    \label{fig:threshold_dist}
\end{figure}
In order to use the qubit level code, we need to map continuous-variable errors arising from finite squeezing to qubit level errors. The correction operators and nonuniform squeezing in a macronode lattice can result in multiple GKP qubits described by a correlated covariance matrix. Ref.~\cite{walshe2024linear} introduced a method for decoding the GKP code that takes these correlations into account.
The binning process in this correlated decoding procedure takes a list of homodyne values $\mathbf{x}$ and aims to find a list of binned outcomes $\mathbf{q}$ minimizing the quantity $\left(\mathbf{q}-\mathbf{x}\right)\Sigma^{-1}\left(\mathbf{q}-\mathbf{x}\right)^T$, where $q_i\in\{ \lfloor{x_i} \rfloor ,\lceil{x_i} \rceil \}$ and $\Sigma$ is the covariance matrix. The error probability of this binning can then be given as 
\small
\begin{equation}
    p\left(i,\mathbf{x},\mathbf{q},\Sigma\right) = \frac{\sum_{k\in \mathbb{Z}}\mathrm{exp}\left(-\frac{1}{2}\left(\mathbf{\tilde{q}}^i_{2k+1}-\mathbf{x}\right)\Sigma^{-1}\left(\mathbf{\tilde{q}}^i_{2k+1}-\mathbf{x}\right)^T\right)}{\sum_{k\in \mathbb{Z}}\mathrm{exp}\left(-\frac{1}{2}\left(\mathbf{\tilde{q}}^i_{k}-\mathbf{x}\right)\Sigma^{-1}\left(\mathbf{\tilde{q}}^i_{k}-\mathbf{x}\right)^T\right)},
\end{equation}
\small
where $i$ indexes the homodyne measurement we are calculating the error probability for and $\tilde{q}^i_{2k+1}$ is the $\mathbf{q}$ where the $i^\mathrm{th}$ element is shifted by $k$. This can be thought of as the ratio of the probability of the homodyne measurement being a result of a displacement by an odd number of teeth (corresponding to a logical error) normalized to the total probability density function for all teeth. We note that the above expression requires the homodyne values and the covariance to be scaled by $1/\sqrt{\pi}$. We use the methodology of correlated decoding in our fault tolerance threshold simulations. 
\subsection{Threshold simulation results}
Fault tolerance thresholds are typically determined as a function of the effective GKP squeezing \cite{Bourassa2021blueprintscalable, larsen2021deterministic, Menicucci2014ft}. It has been shown that with a Gaussian resource of 20 dB and a photon-number resolution of about 40 \cite{takase2021generation}, one can generate GKP states with a high success probability. Reducing the available squeezing naturally decreases this probability \cite{tzitrin2020progress}. Recent works have performed end-to-end simulations to assess the required Gaussian resources \cite{renault2025end,aghaee2025scaling}, which are readily more accessible experimentally \cite{Slusher1985, Yokoyama2013, Vahlbruch2016}. Figure \ref{fig:threshold} shows the logical error rate from our simulations as a function of the cluster squeezing parameter $r$ using GKP distributions from the Monte Carlo runs of GKP generation protocols described earlier. The raw GKP effective squeezing distributions used in the simulations are shown in Fig. \ref{fig:PhANTM_Breedingdist}. For each value of $r$, the logical rate is evaluated at three different code distances $d=3,5,7$. 
$10^5$ tries are performed to estimate each logical error rate to deduce a correction threshold at $r_{th}=11.53$ dB. This value must be compared with previous work, where, with the utilization of 10 PhANTM steps for cat state generation, a cluster squeezing threshold of $r=12.9$ dB was obtained \cite{renault2025end}, leading to an overall improvement of $1.4$ dB. 
Recently, it was demonstrated in Ref. \cite{aghaee2025scaling} that using a 12 dB Gaussian source allows for the generation of GKP states suitable for error correction. In that work, the lack of determinism in state generation was compensated by the use of optical switches that preselect cat states prior to the breeding process. Switches are commonly employed to compensate for the low success probability of GKP state generation, but they come with high losses compared with passive components \cite{chen2023switchreview, psiquantum2025manufacturable}. 

The closest measured point to the error correction threshold is $r=11.5$ dB (see Fig. \ref{fig:threshold}), for which the sum of the effective squeezing in the $q$ and $p$ quadratures is $19.72$ dB (see Fig. \ref{fig:gkp_result}). 
This value is slightly above the symmetric squeezing value of $19.4$ dB \cite{aghaee2025scaling}. To answer whether GKP effective squeezing stochasticity is responsible for the difference, we ran QEC simulations sampling the GKP effective squeezings for each mode from a Gaussian distribution with a standard deviation $\sigma$ centered around different average effective squeezing values. Fig. \ref{fig:threshold_dist} shows the boundary between the correctable and uncorrectable regions for QEC as a function of the GKP effective squeezing in the two quadratures for different values of $\sigma$. As $\sigma$ increases, the boundary shifts toward higher squeezing values, effectively raising the error-correction threshold. The blue marker indicates the average effective squeezing at $11.5$ dB with standard deviations of $1.11$ dB and $0.7$ dB for the effective squeezing in the $q$ and $p$ quadratures, respectively. While the effective squeezing distributions of the Monte Carlo data are not perfectly Gaussian (see \ref{fig:PhANTM_Breedingdist}), the comparison against Gaussian simulated data shows two things. First, realistic stochasticity in GKP effective squeezing introduced due to the passive nature of our architecture changes the threshold marginally from the case where $\sigma$ is zero.  We believe this slight increase in the required squeezing gives a favorable trade-off for bypassing the hardware challenges of on-chip photonic switches for quantum applications \cite{chen2023switchreview, psiquantum2025manufacturable}. Second, knowing the standard deviation in the most stochastic quadrature can help simulate the error correction threshold with Gaussian distributions to gain an upper-bound on the error correction threshold, which may help bypass compute intensive simulations in the future.

\section{Conclusion and Discussion}

We have presented an architecture for fault-tolerant continuous variable quantum computation that achieves a Gaussian error correction threshold of $r = 11.5$ dB cluster squeezing using the RHG surface code. This represents a 1.4 dB improvement over our previous publication~\cite{renault2025end}, bringing the required squeezing parameter closer to experimental feasibility. For comparison, r = 11.5 dB corresponds to $r_0 = 14.5$ dB of two-mode squeezing, while current experimental records stand at $15$ dB in free space \cite{Vahlbruch2016} and $8$ dB in integrated photonics \cite{kashiwazaki2023over, larsen2025integrated}.
The key innovations enabling this threshold reduction include dynamic squeezing protocols integrated with PhANTM on time-multiplexed cluster states, input reset mechanism to prevent cascaded failures in PhANTM as well as numerical optimization of hardware modeling parameters, adaptive breeding with improved rescaling of cat states using teleportation-based squeezing gates, and fewer teleportation operations through targeted cluster engineering. Importantly, our architecture maintains its passive nature, requiring no active switching while achieving photon number resolution requirements of fewer than 10 photons per detector— within the capabilities of superconducting nanowire single-photon detectors\cite{Goltsman2001,marsili2013detecting} and future segmented single-photon avalanche diode arrays \cite{nehra2020photon}. Our simulations demonstrate that the stochasticity in GKP effective squeezing across the two quadratures introduces a small ($\approx 0.15$ dB) shift in the fault tolerance threshold over GKP states with deterministic squeezing. This suggests that targeting deterministic, perfectly balanced GKP states through active switching and post-selection may be unnecessary for realistic fault-tolerant photonic quantum computers. Understanding the interplay between quadrature imbalance and specific error correction codes can further reveal architectural optimization opportunities.

Important extensions of this work will be to refine the protocols qualitatively and develop more realistic numerical simulations that will likely improve the GKP effective squeezing (e.g., by systematically eliminating the effects of post-selection). Another line of work will be to incorporate realistic loss models to quantitatively ascertain the resilience of the architecture to optical losses. Recent work has begun incorporating loss in the context of Gaussian boson sampling-based GKP generation methods \cite{aghaee2025scaling}. For our architecture,
previous studies have examined the effects of post-selection, as well as loss in PhANTM \cite{Eaton2022Phantm} and breeding \cite{renault2025end,solodovnikova2025loss}, but an end-to-end analysis accounting for optical losses in waveguides and beam splitters, detection inefficiencies in homodyne and PNR detectors, and loss-induced reduction of Wigner negativity will be needed for a fuller picture of the advantages of our architecture. 

This work represents an important step toward practical fault-tolerant photonic quantum computing, bringing the required resource thresholds within reach of experimental capabilities. The 11.5 dB error correction threshold for cluster squeezing in a passive architecture requiring only modest photon number resolution and Gaussian resource states is state-of-the-art to the best of our knowledge. 
\section*{Acknowledgments}
We are grateful to our investors, collaborators, and funding agencies for their support.

\section*{Data availability}
The data that support the findings of this paper are
openly available on Harvard Dataverse \cite{Renault2025Dataset}.

\bibliography{biblio}
\newpage
\section{Supplementary Information}
\subsection{Cluster engineering for PhANTM application}\label{sec:appB}

\begin{figure}
    \centering
    \includegraphics[width=0.4\textwidth]{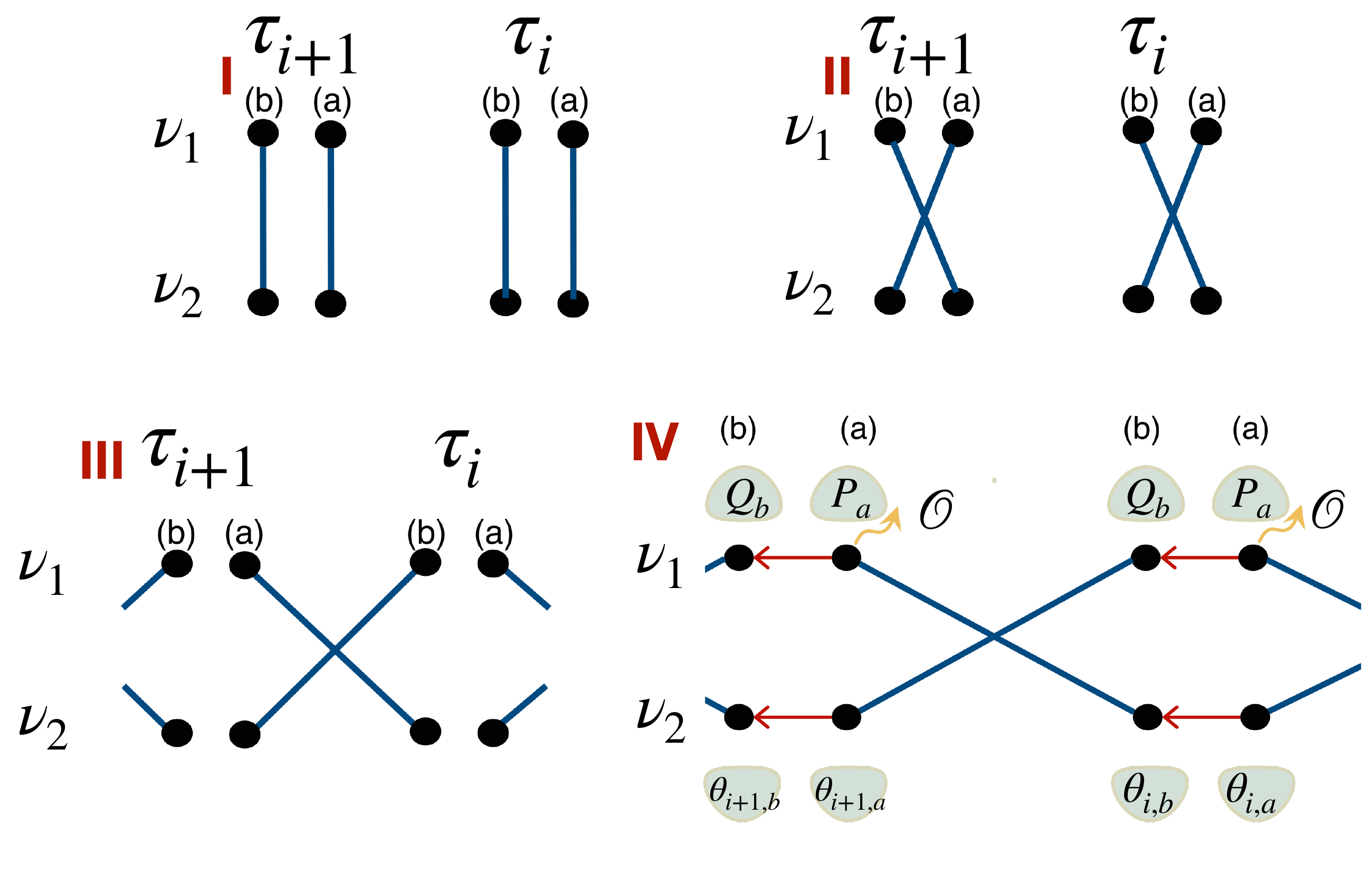} 
    \caption{Schematic of cluster engineering process to reach dual rail frequency-temporal cluster state presented in Fig. \ref{fig:cat_generation}a.   }
    \label{fig:PhANTMcluster}
\end{figure}

Figure~\ref{fig:PhANTMcluster} illustrates the construction of the cluster state used to implement both the PhANTM protocol and the teleportation-based anti-squeezing gate. We consider an on-chip OPO cavity pumped by two identical counter-propagating beams, which generates pairs of two-mode squeezed states (I). These states propagate along separate waveguides, labeled (a) and (b), and their corresponding frequency modes are denoted $\nu_1$ and $\nu_2$. The two frequency modes can be split using a Mach-Zehnder interferometer. Subsequently, $\nu_1$ and $\nu_2$, originating from (a) and (b), respectively, are swapped and recombined into a single waveguide (II). The mode in (a) is then delayed and interferes on a beam splitter with the mode from (b) from the next time step (III). In the circuit shown in Fig.\ref{fig:cat_generation}.b, PhANTM is applied to the frequency mode $\nu_1$, which is subsequently teleported to the mode $\nu_2$ of the next time step, where the anti-squeezing gate is implemented (IV). The total number of photons subtracted across successive PhANTM steps is tracked throughout the protocol. Once this number exceeds a predefined threshold $\mathcal{T}_{\mathrm{ph}}$, the anti-squeezing parameter is adjusted by updating the measurement angles in the homodyne detection.

\subsection{Simulation details in GKP state generation}\label{sec:appA}
\begin{figure}
    \centering
    \includegraphics[width=0.45\textwidth]{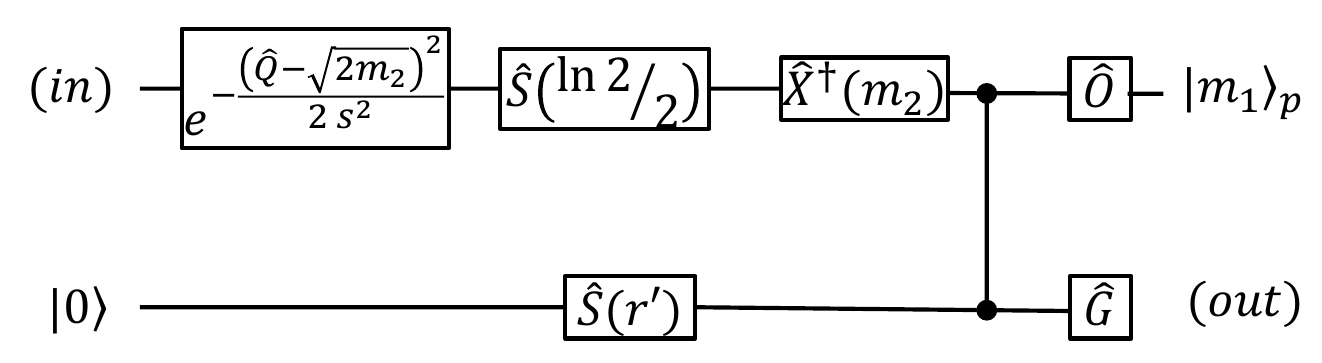} 
    \caption{PhANTM two modes circuit used for Monte Carlo simulation, where $\hat{O}$ stands for multi-photons subtraction, $r'=r\tanh(r_0)/\sqrt{2}$ and $G = \hat{R}^\dagger(\pi/2) \hat{Z}^\dagger(\frac{m_2 \tanh{2r_0}}{\sqrt{2}}) \hat{S}^\dagger (\ln{\frac{\tanh{2r_0}}{\sqrt{2}}})$  }
    \label{fig:Ph_2modes}
\end{figure}

Fig.~\ref{fig:cat_generation} illustrates the circuit used for the Monte Carlo simulation of the PhANTM protocol. However, this full circuit is not computationally optimal. As shown in Ref.~\cite{Eaton2022Phantm}, it can be reduced to an equivalent, more efficient two-mode circuit, which we adopt in our simulations (see Fig.~\ref{fig:Ph_2modes}). For each Monte Carlo run, we simulate 10 successive applications of the PhANTM channel with a given initial squeezing parameter $r_0$ and generate 1000 random samples. In approximately 95\% of the cases, the resulting quantum state can be accurately fit to a squeezed cat state with a fidelity above 0.9. The remaining 5\% of samples fall below this threshold, mainly due to the finite-dimensional Fock space used in the simulation, which is truncated at a cutoff of 60 photons. These points are discarded leading to a conservative lower bound for the amplitudes of cat states possible from our protocols. The simulation code is available upon request. 

To simulate the circuit with a feasible run time and avoid the application of additional displacement corrections, we post-select homodyne measurement outcomes to zero. As a consequence of this post-selection, the resulting cat states have balanced weights. In cases where the homodyne outcome $m_2$ is nonzero, the cat states exhibit unbalanced weights. However, it has been shown that in the subsequent PhANTM step, the likelihood of selecting an $m_2$ value that restores balance is higher than that of worsening the imbalance \cite{Eaton2022Phantm}. Preliminary analyses indicate that breeding unbalanced cat states reduces the effective squeezing of the resulting GKP states. This is countered by two other simulation effects. First, our estimates for cat state amplitudes are conservatively upper-bounded by the simulation Fock dimension. Second, post-selecting homodyne measurements in breeding also provides a conservative estimate of the GKP effective squeezing \cite{Weigand2018}. For these reasons, we expect the fault tolerant threshold to marginally change, perhaps even improve, after incorporating probabilistic homodyne detection in PhANTM and adaptive breeding, along with a higher Fock dimension in our simulations. 

\begin{figure}
    \centering
    \includegraphics[width=0.5\textwidth]{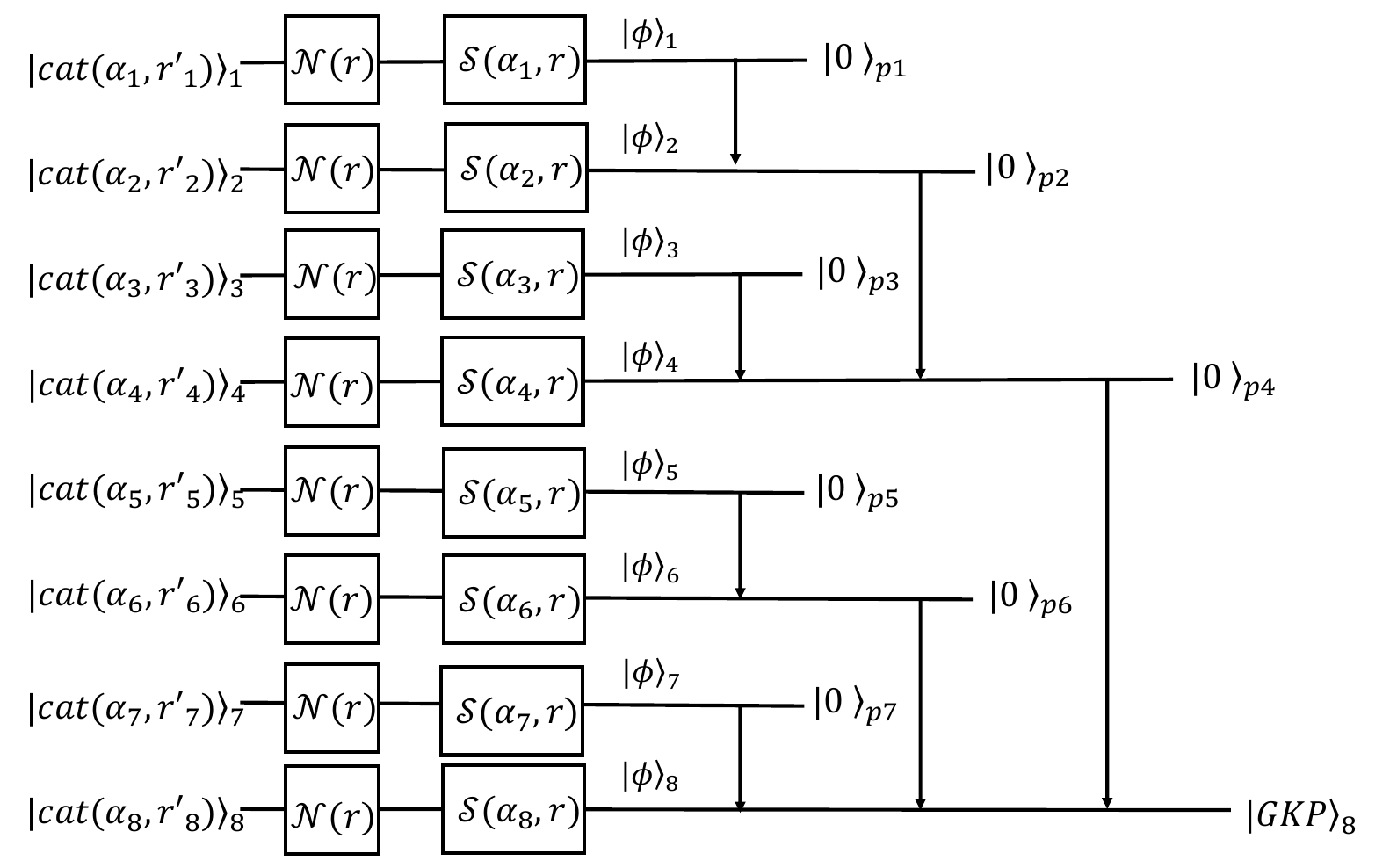} 
    \caption{Adaptive breeding circuit applied in the simulation. As function of the squeezing parameter $r_i$, the state $\ket{\phi}_i$ can be a rescaled cat state $\ket{cat(\alpha,r_i')}$ or a momentum squeezed state $S(r)\ket{0}$.} 
    \label{fig:Breedingcircuit}
\end{figure}

\begin{figure}
    \centering
    \includegraphics[width=0.5\textwidth]{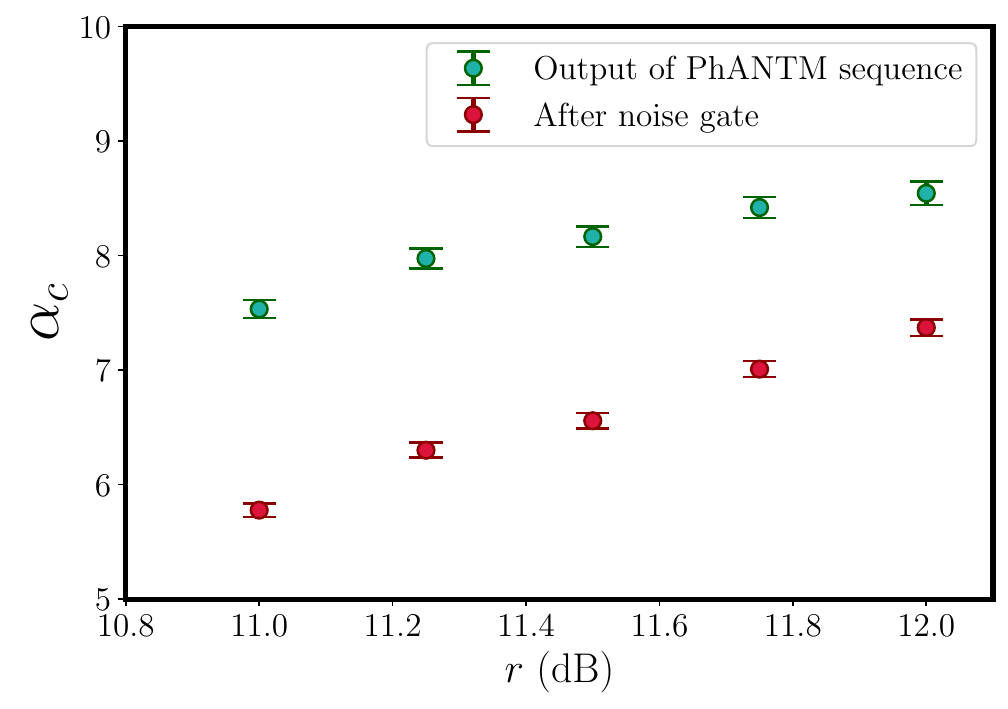} 
    \caption{$\alpha_c$ as a function of cluster squeezing when the cat state is taken directly after PhANTM (10 steps) or the noise gate $\mathcal{N}$. The noise gate simulate the noise due the the squeezing gate used to rescale the cat state before adaptive breeding and applied after PhANTM. The green dot are already presented in the article in Fig. \ref{fig:cat_generation}.b
    }
    \label{fig:noisegate}
\end{figure}

Once the cat states are generated, they serve as the input state for the adaptive breeding circuit shown in Fig. \ref{fig:Breedingcircuit}. Since the protocol uses three breeding steps, eight cat states are required to obtain one GKP state. In the integrated architecture, the cat states are sent through a final squeezing gate, which rescales their amplitude or replaces them with momentum squeezed states by changing the homodyne angles appropriately. Fig. \ref{fig:noisegate} shows the effect of damping on the cat states after PhANTM when going through a teleportation-based squeezing gate with the squeezing set to zero. We see that the size of the corrected cat state ($\alpha_c$) is damped due to the noise channel $\mathcal{N}$, with the effect diminishing at higher values of cluster squeezing $r$. For $r\approx 11.5$ dB and $\Delta \alpha_c\approx2 $, the effect of damping is not negligible.

\begin{figure*}
    \centering
    \includegraphics[width=0.5\textwidth]{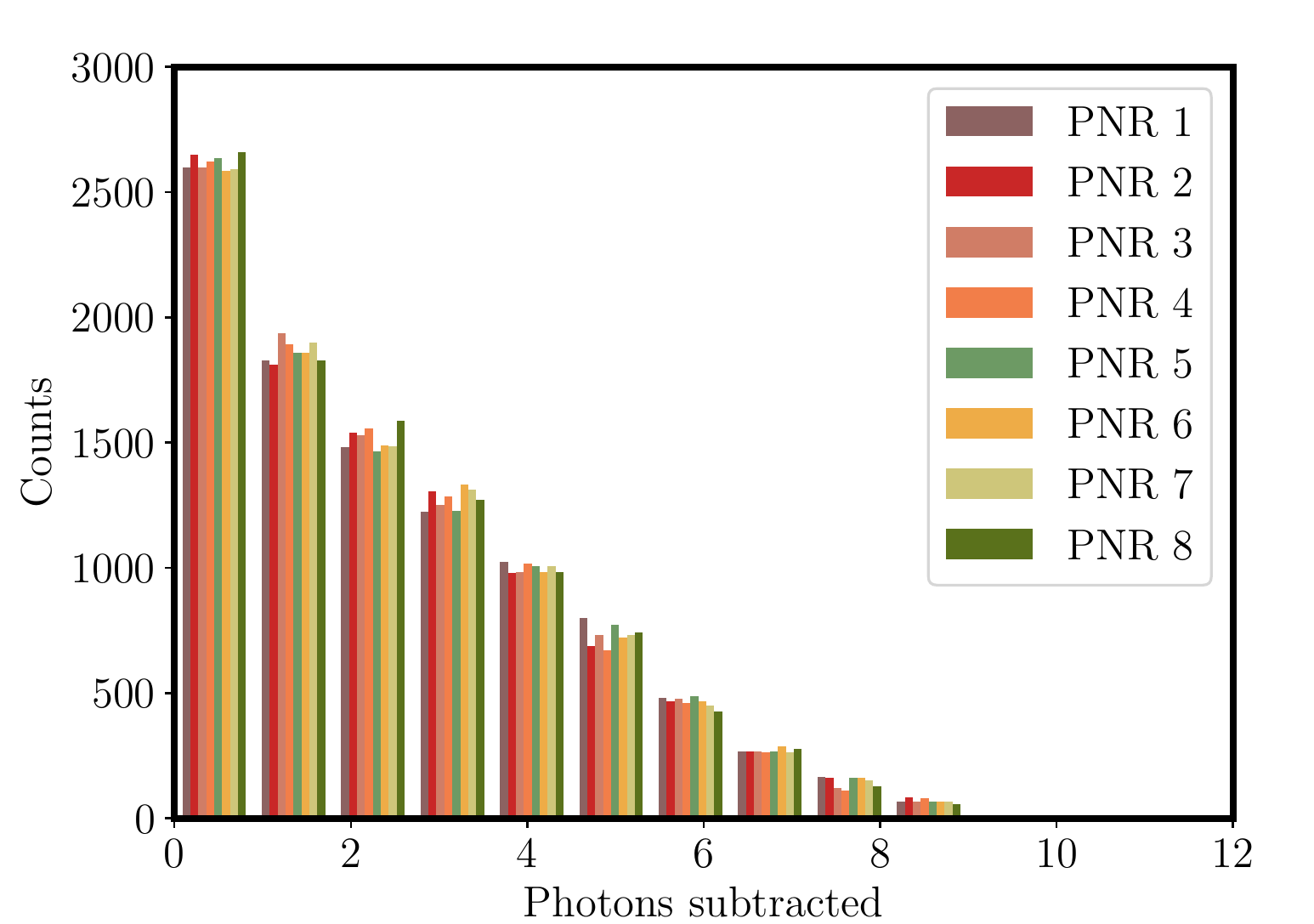} 
    \caption{Distribution of photon subtracted for each PNR detector for $r=12$ dB. Detectors are labeled form 1 to 8.}
    \label{fig:photon_dist}
\end{figure*}

\begin{figure*}
    \centering
    \includegraphics[width=1\textwidth]{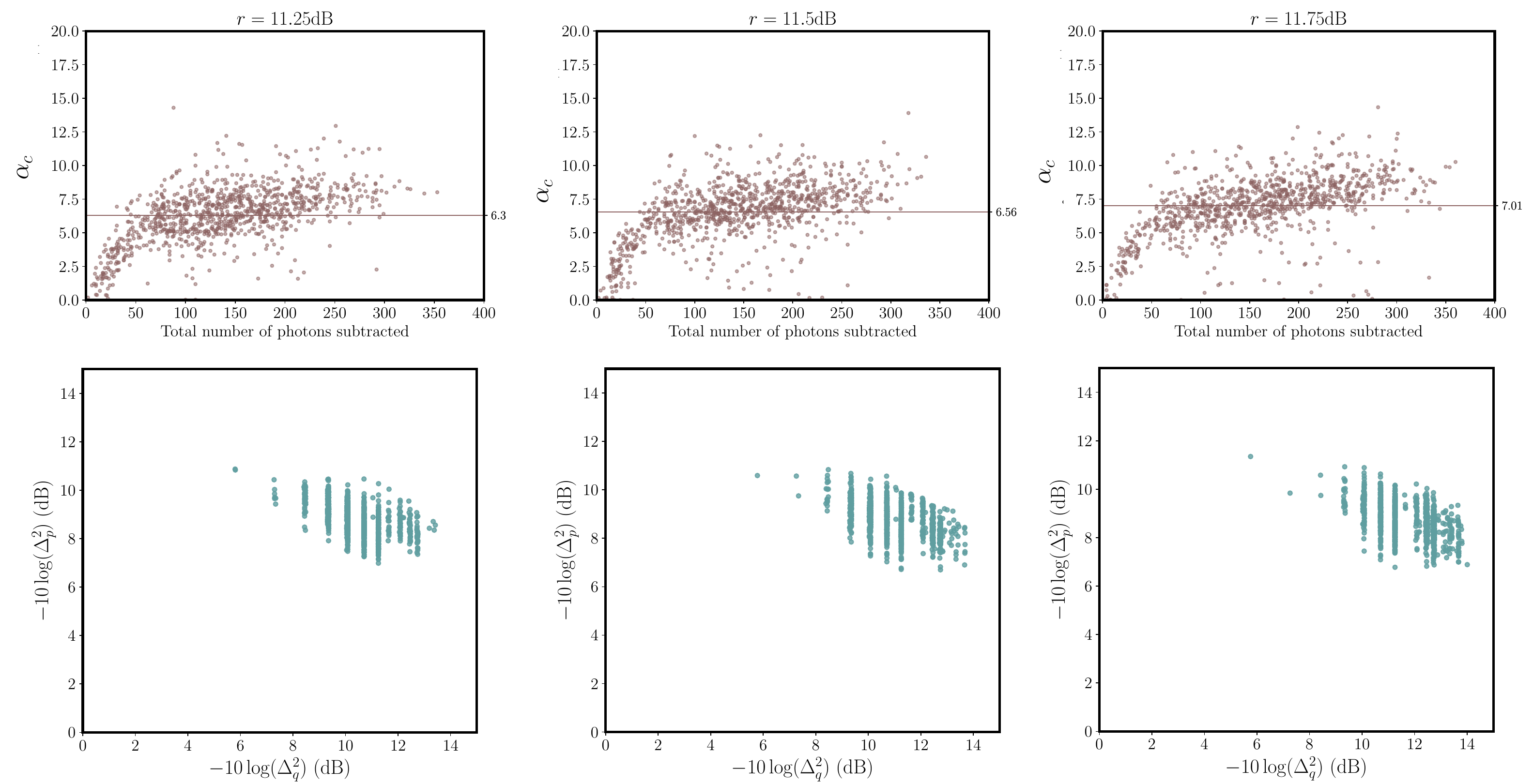} 
    \caption{$\alpha_c$ as a function of the total number of photon subtracted after PhANTM sequence (10 steps) (\textit{top}) and distribution of GKP effective squeezing in $p$ and $q$ quadrature after adaptive breeding (\textit{bottom}) for different $r=11.25$ dB (\textit{left}), $r=11.5$ dB (\textit{center}) and $r=11.75$ dB (\textit{right}).}
    \label{fig:PhANTM_Breedingdist}
\end{figure*}

Fig. \ref{fig:photon_dist} shows the histogram of the detected photons for each of the 8 PNR detectors within PhANTM. The use of a gradient for the beam splitter angles enables similar photon measurement profiles for each detector. We can see that a resolution of less than 10 photons is required by this protocol.
Fig.\ref{fig:PhANTM_Breedingdist} shows the raw data of $\alpha_c$ and the effective squeezing after PhANTM and adaptive breeding processes for different values of cluster squeezing $r$. One can see that the distribution of $\alpha_c$ can be decomposed into two regimes: an initial phase, where $\alpha_c$ increases with the number of subtracted photons, followed by a saturation plateau due to the finite dimension of the Fock space used for the Monte Carlo simulations. For $r=11.75$ dB more data points are located in the plateau compared with $r=11.25$ dB, which makes the average $\alpha_c$ higher, but also likely underestimates the expected amplitude due to constraints on the Fock dimension. The raw data for GKP states show several streaks in the distribution of effective squeezing. These are due to the stochastic set of input cat states. Some of the cat states are replaced by squeezed vacuum states, which affects the effective squeezing in the $q$ quadrature. 

\end{document}